\documentstyle[aps,prl,array,multicol]{revtex}
\def\beginwide{
        \end{multicols} \vspace*{-0.5cm} \noindent
        \rule{3.5in}{.1mm}\rule{.1mm}{5mm} \widetext \medskip } 

\def\endwide{
        \hspace*{3.5in}~\rule[-5mm]{.1mm}{5mm}\rule{3.5in}{.1mm}
        \begin{multicols}{2} \vspace*{-1.0cm} \noindent }   
\begin{document}
\title{On the generalized Landau-Zener problem.}
\author{V.L. Pokrovsky$^{1,2}$, N.A. Sinitsyn$^1$}
\address{$^1$Department of Physics, Texas A\&M University, College Station, \\
Texas 77843-4242,\\
$^2$Landau Institute of Theoretical Physics, Chernogolovka, Moscow region\\
142432, Russia}
\maketitle
\date{}

\begin{abstract}
During the adiabatic time evolution levels crossing violates the
adiabaticity and makes transitions between levels possible. Conventionally
only two energy levels cross simultaneously. The transition probabilities
for this case were found by Landau and Zener \cite{landau},\cite{zener}.
However, the multilevel crossing happens systematically rather than
occasionally if the Hamiltonian possesses a special symmetry. The simplest
physical realization of the multilevel crossing are the Zeeman multiplet in
a varying magnetic field and an electron in one-dimensional chain driven by
the time-dependent electric field. We present asymptotics of the transition
amplitudes for these kinds of the multilevel crossing. They are based on an
exact solution for a model n-state, time-dependent Schr\"{o}dinger equation.
\end{abstract}
\pacs{}
\begin{multicols}{2}
%\pagenumbering{arabic}

%\pagenumbering{arabic}

%\section{Introduction}

Landau-Zener theory is one of the most important and influential results in
non-stationary quantum mechanics. It has numerous applications ranging from
the molecular dissociation \cite{LL}, slow atomic and molecular collisions 
\cite{collisions} to electron transfer in biomolecules \cite{biomolecules}.
Recently it was successfully employed for the experimental determination of
very small tunnel splittings \cite{WS1} and quantum phase interference \cite
{WS2} in molecular nanomagnets.

Landau-Zener (LZ) theory \cite{landau},\cite{zener} deals with an adiabatic
process in a quantum system with discrete spectrum under external bias. The
adiabaticity is violated when a pair of levels move towards each other
strongly enhancing the transitions between the two states. LZ determined the
transition probabilities for the two-level crossing.

LZ theory considers the crossing of more than two levels at one moment of
time to be an unlikely process. However, in some systems such a crossing may
occur systematically stemming from high symmetry of the underlying
Hamiltonian. A realistic example is a multiplet of atomic electronic states
with the total spin $S$ or total rotational moment $J$ larger than 1/2 in
varying external magnetic or electric field. The Zeeman splitting between $%
2S+1$ or $2J+1$ levels regularly vanishes at nodes of the magnetic field.

Another example is provided by localized electronic states in a symmetric
crystal environment. This degeneracy can be lifted by driving external
fields, as in the previous example and by the Jahn-Teller effect. A similar
degeneracy is displayed by a large spin placed in a crystal environment of
high symmetry \cite{vk}. For some specific Hamiltonians employed in
nanomagnets theory the existence of multilevel crossing at specific values
of parameters (components of magnetic field) was shown \cite{diab},\cite
{diab2}. Though this degeneracy probably is unstable with respect to
perturbations of the Hamiltonian, these perturbations are small for real
nanomagnets as it was demonstrated in the experiment by Wernsdorfer {\it et
al.}\cite{WS2}.

In this work we find analytically the transition amplitudes for special
cases of multilevel crossing. For this special symmetry the transition
amplitudes can be expressed in terms of a two-level problem. Physically this
case corresponds to transitions in the Zeeman multiplet placed into a
strong, time-dependent magnetic field directed along a constant axis ($z$)
and a weak, slowly varying or constant magnetic field in the perpendicular
direction ($x$).

The time-dependent Shr\"odinger equations for two levels are:

\begin{eqnarray}
i\dot{b}_1 = E_1(t)b_1+\Delta(t)b_2  \nonumber \\
i\dot{b}_2= E_2(t)b_2 +\Delta(t)b_1  \label{2c2}
\end{eqnarray}

Near a crossing point the dependence of energy levels on time is
approximately linear $E_{\alpha}(t)=\dot{E}_{\alpha}t;\,\,\alpha =1,2$,
whereas the non-diagonal matrix elements of the Hamiltonian can be taken
constants. In terms of new amplitudes $a_{1,2}=e^{-i(\dot{E}_1+\dot{E}_2)t^2
/4}b_{1,2}$, after a time rescaling, equations (\ref{2c2}) can be simplified
as follows \cite{zener}:

\begin{eqnarray}
i\dot{a}_{1} &=&ta_{1}/2+\gamma a_{2}  \nonumber \\
i\dot{a}_{2} &=&-ta_{2}/2+\gamma a_{1}  \label{2c1}
\end{eqnarray}
where $\gamma =\Delta /\sqrt{\dot{\Omega}},\,\,\,\dot{\Omega}=\dot{E}_{1}-%
\dot{E}_{2}$. Eliminating $a_{2}$ from these equations, we find the
parabolic cylinder equation for $a_{1}(t)$. Its solution which has
asymptotics $a_{1}\simeq \exp (-\frac{it^{2}}{4}-i\gamma ^{2}\ln |t|)$ (and $%
a_{2}=0)$ at $t\rightarrow -\infty $ is the Weber function $D_{-i\gamma
^{2}}(e^{i\pi /4}t)$ whose asymptotics are well known \cite{gradstein}. The
scattering matrix for the two-level system is conveniently written in terms
of modified amplitudes $c_{1}=a_{1}\exp (if);\,\,c_{2}=a_{2}\exp (-if)$
where $f=\frac{t^{2}}{4}+\gamma ^{2}\ln |t|$. It reads:

\begin{equation}
U_{\infty }=\left( 
\begin{array}{ll}
\exp (-\pi \gamma ^{2}) & -\frac{\sqrt{2\pi }\exp (-\frac{\pi \gamma ^{2}}{2}%
+\frac{i\pi}{4})}{\gamma \Gamma (-i\gamma ^{2})} \\ 
\frac{\sqrt{2\pi }\exp (\frac{\pi \gamma ^{2}}{2}-\frac{i\pi}{4})}{\gamma
\Gamma ( i\gamma ^{2})} & \exp (-\pi \gamma ^{2})
\end{array}
\right)  \label{2l-S-matr}
\end{equation}

Let us consider a system with the total spin $S>1/2$ in constant magnetic
field $H_{x}$ along $x$-direction and varying with time, much larger in
average field $H_{z}(t)$ along $z$-direction. Its time evolution is
regulated by the Hamiltonian: 
\begin{equation}
\hat{H}_{S}=-h_{x}\hat{S}_{x}-h_{z}(t)\hat{S}_{z}  \label{h1}
\end{equation}
where $h_{\alpha }=g\mu _{B}H_{\alpha };\,\,\alpha =x,z$ and $\hat{S_{z}},%
\hat{S_{x}}$ are the spin operators. In he vicinity of its node $h_{z}(t)$
can be approximated by a linear function $h_{z}(t)=\dot{h}_{z}t$ where $\dot{%
h}_{z}$ is the time derivative of $h_{z}(t)$ taken at the node. After a
proper rescaling of time and energy the Hamiltonian (\ref{h1}) takes a
following form: 
\begin{equation}
\hat{H}=2\gamma \hat{S_{x}}+t\hat{S_{z}}  \label{h2}
\end{equation}
It depends on one dimensionless parameter $\gamma =\frac{h_{x}}{\sqrt{\dot{h}%
_{z}}}$ (the LZ parameter). This Hamilton operator belongs to the $SO(3)$
algebra. This fact allows to derive the evolution of an arbitrary spin $S$
in a varying magnetic field from the solution of Schr\"{o}dinger equation
for spin $1/2$ in the same field. The corresponding evolution operator $%
U_{S}(t)$ is an operator of rotation belonging to the group $SO(3)$ and
acting in its irreducible representation. Since the composition law does not
depend on a specific representation, the resulting evolution operator at a
fixed moment of time represents the same rotation for any spin. Thus, the
problem is reduced to expression of the rotation operator for spin $S$ if it
is known for spin $1/2$. Note that the Hamiltonians (\ref{h1},\ref{h2}) are
most general for the case of 2-level crossing.

The multi-spinor technique is most appropriate for this purpose (see \cite
{LL}, ch. VIII). In general the spin S state can be represented as a direct
symmetric product of 2S spin 1/2 states:
\begin{eqnarray}
\left| S,m\right\rangle =\sqrt{\frac{(S+m)!(S-m)!}{(2S)!}}\left( \left|
++...+--...-\right\rangle +\right.\nonumber\\
\left.\left| ++...-+-...-\right\rangle +...\right)
\label{multispin}
\end{eqnarray}
where each ket contains $S+m$ spins up and $S-m$ spins down and all
permutations are performed. Let the $SU(2)$ matrix rotating spin 1/2 states
is:

\begin{equation}
u=\left( 
\begin{array}{ll}
a & b \\ 
-b^{*} & a^{*}
\end{array}
\right)  \label{su2}
\end{equation}
with $|a|^{2}+|b|^{2}=1$. Equivalently an individual spinor is transformed
according to:

\begin{eqnarray}
\left| +\right\rangle &\rightarrow &a\left| +\right\rangle -b^{*}\left|
-\right\rangle ;  \nonumber \\
\left| -\right\rangle &\rightarrow &b\left| +\right\rangle +a^{*}\left|
-\right\rangle  \label{transf2}
\end{eqnarray}
The transformation for the state (\ref{multispin}) can be obtained as the
direct product of transformations (\ref{transf2}).
\beginwide
\begin{eqnarray}
\left| S,m\right\rangle \rightarrow \sqrt{\frac{(S+m)!(S-m)!}{(2S)!}}%
a^{S+m}(-b^{*})^{S-m}\left| S,S\right\rangle +\nonumber\\
\left( \frac{\sqrt{2S}(2S-1)!}{%
(S-m-1)!(S+m)!}a^{S+m}(-b^{*})^{S-m-1}a^{*}+
 \frac{\sqrt{2S}(2S-1)!}{(S+m-1)!(S-m)!}a^{S+m-1}(-b^{*})^{S-m}b\right)
\left| S,S-1\right\rangle ...  \nonumber
\end{eqnarray}
\endwide
A general matrix element of the rotation operator $\langle m\mid U_{S}\mid m^{\prime }\rangle$ for the spin S representation  is expressed
in terms of $a,b,a^{*},b^{*}$ in the following way \cite{edmonds}, \cite{3v}%
: 
\beginwide
\begin{equation}
\langle m\mid U_{S}\mid m^{\prime }\rangle =\left[ \frac{(S+m^{\prime
})!(S-m^{\prime })!}{(S+m)!(S-m)!}\right] ^{1/2}a^{m^{\prime
}+m}b^{m^{\prime }-m}P_{S-m^{\prime }}^{m^{\prime }-m,m^{\prime
}+m}(2|a|^{2}-1)  \label{general}
\end{equation}
\endwide
where $P_{n}^{a,b}(x)$ are the Jacobi polynomials \cite{erdelyi}. The matrix
elements possess the following symmetry properties: $\langle -m\mid
U_{S}\mid -m^{\prime }\rangle =(-1)^{|m|+|m^{\prime }|}\langle m\mid
U_{S}\mid m^{\prime }\rangle ^{*}$, $\left| \langle m\mid U_{S}\mid
m^{\prime }\rangle \right| =\left| \langle m^{\prime }\mid U_{S}\mid
m\rangle \right| =\left| \langle -m^{\prime }\mid U_{S}\mid -m\rangle
\right| $.

Equation (\ref{general}) displays oscillations of matrix elements associated
with oscillatory behavior of the Jacobi polynomials. The number of nodes $%
N(m,m^{\prime})$ of the matrix elements $\langle m\mid U_S\mid
m^{\prime}\rangle$ can be determined geometrically as the number of the
square shell to which it belongs in the square matrix. We accept the number
of the external square with $max(|m|,|m^{\prime}|)=S$ for zero and number is
increasing when the square shell size is decreasing. Analytically $%
N(m,m^{\prime})=S-max(|m|,|m^{\prime}|)$. Due to symmetry several matrix
elements (2, 4 or 8) become zero at the same value of $|a|$ or $\gamma$.
Central matrix elements have maximal number of nodes ($S$ for integer spins, 
$S-1/2$ for half-integer spins).

The scattering matrix for our problem can be found from the general
expression (\ref{general}) by substitution: 
\begin{equation}
a=\exp (-\pi \gamma ^{2}),\,\,\,\,b=-\frac{\sqrt{2\pi }\exp {(\frac{\pi
\gamma ^{2}}{2}+\frac{i\pi }{4})}}{\gamma \Gamma (-i\gamma ^{2})}
\label{a,b}
\end{equation}
which follows from comparison of equations (\ref{2l-S-matr},\ref{su2}). To
make our results more visual, we present explicitly the scattering matrices
for spins $S=1$ and $S=3/2$:

\noindent $S=1$.

\begin{equation}
U_{1}=\left( 
\begin{array}{lll}
a^{2} & \sqrt{2}ab & b^{2} \\ 
-\sqrt{2}ab^{*} & 2|a|^{2}-1 & \sqrt{2}a^{*}b \\ 
b^{*2} & -\sqrt{2}a^{*}b^{*} & a^{*2}
\end{array}
\right)  \label{U11}
\end{equation}

$S=3/2.$

\begin{equation}
U_{3/2}=\left( 
\begin{array}{llll}
a^{3} & \sqrt{3}a^{2}b & \sqrt{3}ab^{2} & b^{3} \\ 
-\sqrt{3}a^{2}b^{*} & (3|a|^{2}-2)a & (3|a|^{2}-1)b & \sqrt{3}a^{*}b^{2} \\ 
\sqrt{3}ab^{*2} & -(3|a|^{2}-1)b^{*} & (3|a|^{2}-2)a^{*} & \sqrt{3}a^{*2}b
\\ 
-b^{*3} & \sqrt{3}a^{*}b^{*2} & -\sqrt{3}a^{*2}b^{*} & a^{*3}
\end{array}
\right)  \label{U3/2}
\end{equation}
To save space we present only several transition probabilities for $S=2$, 
all the rest can be found from symmetry properties (see above):
$W_{2,2-k}=\frac{4!}{k!(4-k)!}(|a|^2)^{4-k}(1-|a|^2)^k,\,\,k=0...4;
W_{1,1}=|a|^4(4|a|^2-3)^2; W_{1,0}=6|a|^2(1-|a|^2)^2(2|a|^2-1)^2;
W_{1,-1}=(1-|a|^2)^2(4|a|^2-1)^2; W_{0,0}=(6|a|^4-6|a|^2+1)^2$.
%\begin{equation}
%U_{2}=\left( 
%\begin{array}{lllll}
%a^{4} & 2a^{3}b & \sqrt{6}a^{2}b^{2} & 2ab^{3} & b^{4} \\ 
%-2a^{3}b^{*} & (4|a|^{2}-3)a^{2} & -\sqrt{6}(2|a|^{2}-1)ab^{*} & 
%(4|a|^{2}-1)b^{2} & 2a^{*}b^{3} \\ 
%\sqrt{6}a^{2}b^{*2} & -\sqrt{6}(2|a|^{2}-1)ab^{*} & 6|a|^{4}-6|a|^{2}+1 & 
%\sqrt{6}(2|a|^{2}-1)a^{*}b & \sqrt{6}a^{*2}b^{2} \\ 
%-2ab^{*3} & (4|a|^{2}-1)b^{*2} & -\sqrt{6}(2|a|^{2}-1)a^{*}b^{*} & 
%(4|a|^{2}-3)a^{*2} & 2a^{*3}b \\ 
%b^{*4} & -2a^{*}b^{*3} & \sqrt{6}a^{*2}b^{*2} & -2a^{*3}b^{*} & a^{*4}
%\end{array}
%\right)
%\end{equation}

Additionally we present transition probabilities for $S=1$: 
%\begin{equation}
%U_{1}=\left( 
%\begin{array}{lll}
%\exp (-2\pi \gamma ^{2}) & -\frac{2\sqrt{\pi }\exp (-\frac{3\pi
%\gamma ^{2}}{2}+\frac{i\pi}{4})}{\gamma \Gamma (-i\gamma ^{2})} & \frac{2i\pi
% \exp
%(-\pi \gamma ^{2})}{\gamma ^{2}(\Gamma (-i\gamma ^{2}%
%))^{2}} \\ 
%\frac{2\sqrt{\pi }\exp (-\frac{3\pi \gamma ^{2}}{2}-\frac{i\pi}{4})}{\gamma 
%\Gamma (%
%i\gamma ^{2})} & -1+2\exp (-2\pi \gamma ^{2}) & \frac{-2\sqrt{\pi }\exp
%(-\frac{3\pi \gamma ^{2}}{2}+\frac{i\pi}{4})}{\gamma 
%\Gamma (-i\gamma ^{2})}\\ 
%\frac{2i\pi \exp (-\pi \gamma ^{2})}{\gamma ^{2}(\Gamma (%
%i\gamma ^{2}))^{2}} & \frac{2\sqrt{\pi }\exp (-\frac{3\pi \gamma ^{2}}
%{2-\frac{i\pi}{4}%
%})}{\gamma \Gamma (i\gamma ^{2})} & \exp (-2\pi \gamma ^{2})
%\end{array}
%\right)   \label{U1-re}
%\end{equation}

%\begin{equation}
%W=\left( 
%\begin{array}{lll}
%e^{ -4\pi \gamma ^{2}} & 2(e^{ -2\pi \gamma ^{2}}-e^{ -4\pi \gamma ^{2}}) & 
%\left( 1-e^{ -2\pi \gamma ^{2}}\right) ^{2} \\ 
%2(e^{ -2\pi \gamma ^{2}}-e^{ -4\pi \gamma ^{2}}) & \left( 1-2e^{ -2\pi
%\gamma ^{2}}\right) ^{2} & 2(e^{ -2\pi \gamma ^{2}}-e^{ -4\pi \gamma ^{2}})
%\\ 
%\left( 1-e^{ -2\pi \gamma ^{2}}\right) ^{2} & 2(e^{ -2\pi \gamma ^{2}}-e^{
%-4\pi \gamma ^{2}}) & e^{-4\pi \gamma ^{2}}
%\end{array}
%\right)  \label{prob}
%\end{equation}
\begin{eqnarray}
W_{1,1}=e^{-4\pi\gamma^2};\,\,W_{1,0}=2(e^{ -2\pi \gamma ^{2}}-e^{ -4\pi \gamma ^{2}});\nonumber\\
W_{1,-1}=\left( 1-e^{ -2\pi \gamma ^{2}}\right) ^{2};\,\,W_{0,0}=
\left( 1-2e^{ -2\pi\gamma ^{2}}\right) ^{2}.
\label{prob}
\end{eqnarray}
Other probabilities can be readily found from symmetry relations.
As a consequence of Jacobi polynomials oscillation, the matrix elements of
inner squares have nodes at some special values of $\gamma $. Thus, element $%
(U_{1})_{00}$ is zero at $|a|^{2}=1/2$ or at $\gamma =\sqrt{\frac{1}{2\pi }%
\ln 2}\approx 0.332$. Elements of the matrix $%
(U_{3/2})_{1/2,1/2}=(U_{3/2})_{-1/2,-1/2}^{*}$ become zero at $|a|^{2}=2/3$,
i.e. at $\gamma =\sqrt{\frac{1}{2\pi }\ln \frac{3}{2}}\approx 0.254$. Other
two matrix elements $(U_{3/2})_{1/2,-1/2}=(U_{3/2})_{-1/2,1/2}^{*}$ become
zero at $|a|^{2}=1/3$, i.e. at $\gamma =\sqrt{\frac{1}{2\pi }\ln 3}\approx
0.418$. In the matrix $U_{2}$ 4 independent matrix elements have nodes: $%
(U_{2})_{11}$ at $|a|^{2}=3/4$; $(U_{2})_{00}=0$ at $|a|^{2}=\frac{1}{2}%
\left( 1\pm \frac{1}{\sqrt{3}}\right) $; $(U_{2})_{10}=0$ at $|a|^{2}=1/2$; $%
(U_{2})_{1,-1}=0$ at $|a|^{2}=1/4$.

{\it Electron motion driven by electric field}. Here we present another
exact solution. It describes transitions at intersection of infinite number
of equidistant levels. This system can be described by the Hamiltonian,
which matrix representation is

\begin{equation}
H_{nm}=nh(t)\delta_{nm}+\gamma t_{nm}  \label{h3}
\end{equation}
where $t_{nm}=1$ if $|n-m|=1$ and zero otherwise. The physical
implementation of this model is an electron tunneling in a discrete
equidistant chain placed into a varying homogeneous electric field directed
along the chain. We have introduced a natural set of states $|n>$ located at
the $n$-th site. The tunneling between the sites is suppressed by the
external field since it lifts the degeneration of energy. However, when the
field becomes zero the tunneling rate grows. Near this point we accept the
time dependence of the electric field as linear. Then, after rescaling of
time the initial Hamiltonian (\ref{h3}) reads:

\begin{equation}
H_{mn}=t\delta _{mn}+g(\delta _{m,n+1}+\delta _{m,n-1})  \label{h4}
\end{equation}
where $g=\gamma /\sqrt{eE\Delta }$. It corresponds to a system of
Shr\"{o}dinger equations:

\begin{equation}
i\dot{a}_n=nta_n+g(a_{n-1}+a_{n+1})  \label{z1}
\end{equation}
The problem is to find the scattering matrix for this system. In other
words, we are looking for an asymptotic at $t\rightarrow +\infty$ of a
solution $a_n(t)$ which obeys the initial condition $|a_n(t)|^2=%
\delta_{n,n^{\prime}}$ at $t\rightarrow -\infty$.

Let introduce an auxiliary function $u(\varphi ,t)=\sum_{n=-\infty }^{\infty
}a_{n}(t)e^{in\varphi }$. System (\ref{z1}) is equivalent to a following
equation in partial derivatives for $u(\varphi ,t)$: 
\begin{equation}
\frac{\partial u}{\partial t}+t\frac{\partial u}{\partial \varphi }+2ig\cos
\varphi u\,=\,0  \label{partial}
\end{equation}
We should find a solution of this equation which obeys the initial
condition: $u(\varphi ,t)\rightarrow \exp [in^{\prime }(-\frac{t^{2}}{2}%
+\varphi )]$ at $t\rightarrow -\infty $. Given the solution $u(\varphi ,t)$,
the amplitudes $a_{n}(t)$ can be found by the inverse Fourier
transformation: $a_{n}(t)=\frac{1}{2\pi }\int_{0}^{2\pi }u(\varphi
,t)e^{-in\varphi }d\varphi $. The solution of eq. (\ref{partial}) which
obeys proper boundary conditions is: 
\begin{eqnarray}
u(\varphi ,t)=\exp \left[ -i\left( 2g\int\limits_{-\infty }^{t}\cos \left(
\varphi -\frac{t^{2}}{2}+\frac{t^{^{\prime }2}}{2}\right) dt^{^{\prime
}}+\right.\right.\nonumber\\
\left.\left. n^{^{\prime }}\left( \varphi -\frac{t^{2}}{2}\right) \right) 
\right] 
\label{u}
\end{eqnarray}
Putting $t=+\infty $ in the solution (\ref{u}) and taking the inverse
Fourier-transform, we arrive at the asymptotics: 
\begin{equation}
a_{n}(t)\approx \exp (-int^{2}/2+i(n^{\prime }-n)\pi /4)J_{|n-n^{\prime
}|}(2\sqrt{2\pi }g)  \label{ampl}
\end{equation}
Thus, the scattering amplitudes in terms of modified states with the fast
phase factor $\exp (-int^{2}/2)$ incorporated are: 
\begin{equation}
\langle n\mid U_{\infty }\mid n^{\prime }\rangle =e^{i(n^{\prime }-n)\pi
/4}J_{|n-n^{\prime }|}(2\sqrt{2\pi }g)  \label{scat}
\end{equation}
It displays infinite number of oscillations with the LZ parameter $g$.
However, for large $|n-n^{\prime }|$ the oscillations start with $g\gg
|n-n^{\prime }|$. For smaller values of $g$ the amplitudes are small.

This result can be easily extended to a more general non-diagonal matrix
elements dep in which hopping from any site to any other site is allowed and
its amplitude depends only on distance between sites. The corresponding
Hamiltonian has the following matrix elements:

\begin{equation}
H_{mn}=nt\delta _{mn}+g_{m-n};\quad g_{-k}=g_{k}^{*}  \label{ext}
\end{equation}
For simplicity we present below the result for real $g_{k}=g_{-k}$:

\begin{eqnarray}
\left\langle n\right| U_{\infty }\left| n^{^{\prime }}\right\rangle =\frac{%
e^{i(n^{\prime }-n)\pi /4}}{2\pi }%
\int \limits_{0}^{2\pi }\exp \left( -i2\sqrt{2\pi }f(\varphi
,g_{j})\right.\nonumber\\
\left. +i(n^{^{\prime }}-n)\varphi \right) d\varphi   
\label{amplext}
\end{eqnarray}
where $f(\varphi ,g_{j})=\sum\nolimits_{k}\frac{g_{k}}{\sqrt{k}}\cos
k\varphi $.

In conclusion, we presented a generalization of the LZ theory for arbitrary
number of crossing equidistant levels in the case of time-dependent magnetic
field and an electron in an infinite chain subject to a time-dependent
electric field. In both cases high symmetry of the problem allows to find
not only asymptotics, but also intermediate values of the amplitudes. A
number of analytical solutions exists for spin $1/2$ in time dependent
magnetic fields \cite{td} and all they can be generalized for higher spins.
At the same time, our results are not universal. If the number of levels is
more than two, they can approach to crossing point not equidistantly, and
non-diagonal elements of Hamiltonian do not necessarily depend on one
parameter only. In the majority of
cases when the levels intersect systematically they remain equidistant, but
the non-diagonal matrix elements of the Hamiltonian are different from above
considered. Even in a simple case of a constant electric field perpendicular
to the time dependent magnetic field, the Hamiltonian contains the elements
changing the projection $m$ by $\pm 2$. For this case we have found general
statement on separability and can reduce the problem to already solved one
for $S\leq 2$. On the other hand, our solutions demonstrate phenomena that
are probably universal, like oscillations of the transition probabilities.

\noindent {\it Acknowledgments}. This work was supported by NSF under the
grant DMR 0072115 and by DOE under the grant DE-FG03-96ER45598. One of us
(VP) acknowledges the support of the Humboldt Foundation. We thank A.
Kashuba and W. Saslow for important remarks.

\end{multicols}
\end{document}